\documentclass[12pt]{iopart}
\usepackage{iopams}  
\usepackage{graphicx}
\begin{document}

\title{Competition between structural and intrinsic dispersion in delay through a left handed medium}

\author{Dheeraj Golla$^1$, Subimal Deb$^2$ and S Dutta Gupta$^2$}

\address{$^1$ Indian Institute of Technology Kharagpur, Kharagpur 721302 , India \\ $^{2}$ School of Physics, University of Hyderabad, Hyderabad 500046, India}
\ead{sdghyderabad@gmail.com}
\begin{abstract}
We study resonant tunneling  through a layered medium with a negative index medium (NIM) slab as a constituent layer. We demonstrate large delays in transmission mediated by the surface and the guided modes of the structure with low losses. We show how important it is to include NIM dispersion for correct assessment of the nature and magnitude of the delay. We also point out the role of NIM absorption for the feasibility of such compact delay devices. 
\end{abstract}

\maketitle
\section{Introduction}
In recent years there has been a great deal of interest in negative index materials (NIMs), which can exhibit exotic properties \cite{veselago1968,shalaevbook,shalaev2007,sar}. The rich physics of these materials (not occuring in nature) were discussed theoretically by Veselago \cite{veselago1968}, though their potential application for beating the diffraction limit \cite{pendry2000} and experimental realization in the microwave range \cite{shelby1} triggered off the vast current activities. Now such materials (also known as metamaterials or left handed materials) have been fabricated in other wavelength domains right upto the visible range \cite{shalaev2007,sar}. Applications of these metamaterials now range from super-lensing and super-resolution \cite{pendry2000,fang} to lasing spasers \cite{zheludev} and optical nanocircuits \cite{engheta}; invisibility cloaks 
to electromagnetically induced transparency and slow or stopped light \cite{sdgprb,zheludev-arxiv}, etc. The real challenge now is to fabricate a low loss metamaterial with high figure of merit (FOM=$-Re(n)/Im(n)$ where $n$ is the refractive index of the metamaterial) and to broaden the frequency domain where both the permittivity and permeability are negative. It is also important to push the domain to even higher frequencies maintaining low losses.
\par
In the context of slow light using NIM, the role of dispersion cannot be underestimated. As pointed out by Veselago, a lossless NIM is essentially dispersive \cite{veselago1968}. The important role played by dispersion has been felt by others in the context of waveguides \cite{kivshar2003} and especially with reference to cavity QED applications \cite{zhu1,fleischhauer2005,xu2009}. In the context of any finite structure, the frequency dependence of its important characteristics (e.g., transmission through it) emerges from two sources \cite{sdg1998}. The first is due to the material dispersion of its constituents, while the second is due to the boundary conditions. For example, the transmission through an empty Fabry-Perot (FP) cavity, or the cavity filled with a dispersive material, can be quite different. These characteristics can be quite different as compared to the case of propagation through the bulk dispersive sample. We refer to the first as material or intrinsic dispersion, while the second is labelled as structural dispersion. In this paper our goal is to bring out the salient features of how these sources of dispersion affect the transmission through a NIM guide in a resonant tunnelling (RT) geometry \cite{pendry2008,sdg2009}. Unfortunately, many of the papers on the NIM waveguides do not address the issue of material dispersion for calculating the overall dispersion features \cite{kong2003}. In many others the system is investigated at a particular wavelength, concentrating on structural dispersion only, avoiding thus all material dispersion related issues \cite{ jose}.
\par
Our choice of the RT geometry is motivated by several facts. It was shown recently that light can be slowed down using a gap plasmon guide in RT configuration \cite{sdg2009}. Very recently a multilayered metal/dielectric structure in the same configuration was studied experimentally to show enhanced transmission mediated by the modes of the structure \cite{pendry2008}. We show that a NIM guide in RT configuration can lead to significant delays provided the losses are low. Our calculations are carried out with experimental data for the permittivity and the permeability of the NIM \cite{dolling2006}, albeit with the approximation of low losses. We also assumed a homogeneous and isotropic character for the NIM. We studied both $\sin{}$/$\cos{}$ and the $\sinh{}$/$\cosh{}$ types of modes of the NIM guide \cite{kong2003}. We refer to the former as guided modes and the latter as the plasmon-like modes. Indeed, the latter ones (TE polarized) resemble the TM polarized surface plasmon polaritons of a thin metal film or a gap plasmon guide. We correlate the angular location of the peaks in resonant transmission to the corresponding modes in a bare NIM guide. In all our calculations, along with retention of both the sources of dispersion, we present results where material dispersion is suppressed. We show that such a step can lead to wrong conclusions as regards the nature and the magnitude of the delay. In fact, there is a danger of predicting superluminal transit while actually it is subluminal. It is believed that NIMs have to be essentially lossy \cite{stockman2007, kinsler2008prl}. In order to have some quantitative idea about the effect of absorption on the delay, we assume a causal response of the NIM medium. We show that even in presence of a much improved magnetic response it may be a difficult job to have high-Q guided and surface modes in the optical and near IR domains. Our study thus reiterates the important observation of Veselago in the context of absorption and dispersion of a realistic metamaterial.
\par
The structure of the paper is as follows. In section 2, we consider a generic negative index material slab in order to reveal the roles of intrinsic and structural dispersion. We show that there can be qualitative differences in the group delay if the material dispersion is suppressed. In section 3, we pick the NIM parameters from a recent experiment \cite{dolling2006}, albeit with the assumption of small damping. We study group delay through a structure supporting resonant tunnelling. After a brief discussion of the NIM dispersion and the resulting delay in bulk, we investigate the plasmon-like modes of the structure. We show that the excitation of such modes can lead to enhanced delay as compared to FP modes, albeit at the expense of lower transmission. In section 4, we discuss the guided modes of the same structure, except for lower values of refractive index of the spacer layers. Like in the case of plasmon-like modes, these guided modes are also shown to lead to enhanced delay. In section 5, we fit the experimental data \cite{dolling2006} for the magnetic resonance to a causal response and vary the relevant parameters in order to obtain better features of the NIM. We show that even a substantial improvement of the magnetic response fails to excite true high-Q guided and surface modes in the NIM guide. Finally, we summarize the main results in Conclusions. We make an important observation on how the large damping in the present day metamaterials can destroy the resonant tunnelling features, stressing the dire need to fabricate `transparent'  metamaterials.
\section{Group delay in a FP cavity containing a generic NIM}
Consider first the propagation of an optical pulse through a bulk NIM sample. 
Let the permittivity $\epsilon$ and the permeability $\mu$ of the NIM be given by the following expressions \cite{sdgprb,kroll2000}
\begin{equation}
\epsilon(\omega)=\frac{\omega^2-\omega_p^2}{\omega^2}, ~~ \mu(\omega)=\frac{\omega^2-\omega_b^2}{\omega^2-\omega_0^2+i\Gamma\omega} .\label{eq:1}
\end{equation}
Recall that the group delay for a segment of length $d$ in the bulk is given by
\begin{equation}
\tau = \frac{d}{v_g} = \left( \frac{d}{c} \right)n_g , ~ n_g = n(\omega) + \omega \frac{\partial n}{\partial \omega}, \label{eq:2}
\end{equation}
where $n_g$ and $n$ are the group index and the refractive index of the material, respectively. It can be easily seen from (\ref{eq:2}) that group index $ n_g$ is the same as the normalized delay $\tau/(d/c)$. Since the source of delay in (\ref{eq:2}) is the material dispersion, we will refer to this as delay due to intrinsic dispersion. From a different viewpoint, delay $\tau$ through a segment of length $d$ can be linked to the frequency derivative of the phase $\phi_t$ (accumulated over distance $d$) of the (amplitude) transmission coefficient $t$ ($ = |t|e^{i\phi_t} $ ) \cite{wigner1955} as
\begin{equation}
\tau=\frac{\partial \phi_t}{\partial \omega}|_{\omega=\omega_c},\label{eq:3}
\end{equation}
where $\omega_c$ is the carrier frequency of the pulse. $\tau$ in (\ref{eq:3}) is also referred to as the Wigner phase time.

As mentioned earlier, for finite structures, there is another source of frequency dependence of important parameters like transmission and reflection through it. For example, if one considers a slab of width $d$ of a material with some refractive index, embedded in a medium with another index, the characteristic signatures of the Airy resonances of the FP cavity will be imprinted on the delay features. We will refer to such features (with neglect of material dispersion), as delay due to pure structural dispersion. The overall delay emerges as an interplay and competition of these two distinct sources of dispersion. It is thus clear that none of the sources of dispersion can be ignored for correct assessment of the delay. Irrespective of the nature of the delay, it can be evaluated by calculating the phase of the amplitude transmission coefficient and using the expression for the Wigner phase time.
\begin{figure}[b]
\centering
\includegraphics[width=11cm]{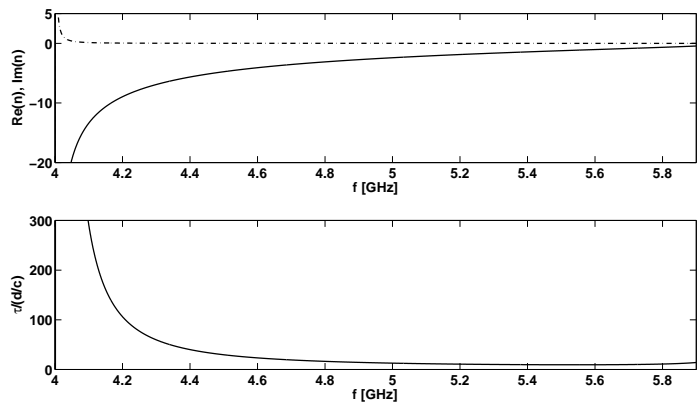}
\caption{(a) Real (solid) and imaginary (dotted) parts of the refractive index of the NIM, (b) the normalized delay for a bulk sample. The parameters chosen are: $f_p = 12 $ GHz, $f_b = 6 $ GHz, $f_0 = 4 $ GHz and $\Gamma/\omega_0 =10^{-3}$.}\label{fig:1}
\end{figure}

The group delay of a pulse passing through a NIM FP cavity was studied in detail and an analytical expression for the delay was derived \cite{sdgprb}. It was shown that such a system can lead to large group delays at frequencies corresponding to the Airy resonances.  A closer inspection of the expression of the Wigner phase time reveals that the delay has a very pronounced dependence on the frequency derivatives of the permittivity and the permeability. It is thus expected that negligence of the material dispersion would lead to quite erroneous conclusions, even to the extent of the sign of the delay. In other words, one may wrongly predict superluminal transit, while it is actually subluminal.

In order to have a quantitative assessment of the contributions of intrinsic and structural dispersion for the delay, we first study the dispersive properties of the bulk material given by (\ref{eq:1}). Parameters were taken from the work of Smith \etal \cite{kroll2000}, namely $f_p = 12 $ GHz, $f_b = 6 $ GHz, $f_0 = 4 $ GHz and $\Gamma/\omega_0 =10^{-3}$, where $f_{p,b,0}=\omega_{p,b,0}/(2\pi)$. The frequency dependence of the real and imaginary parts of the  refractive index is shown in figure \ref{fig:1}a. In the bottom panel (figure \ref{fig:1}b) we have plotted the time taken for propagation through a distance $d$ in the bulk sample. It is clear that over the specified range of frequencies the bulk material exhibits normal dispersion which leads to a delay of the pulse. If propagation is through a slab of such a 
\begin{figure}[t]
\centering
\includegraphics[width=12cm]{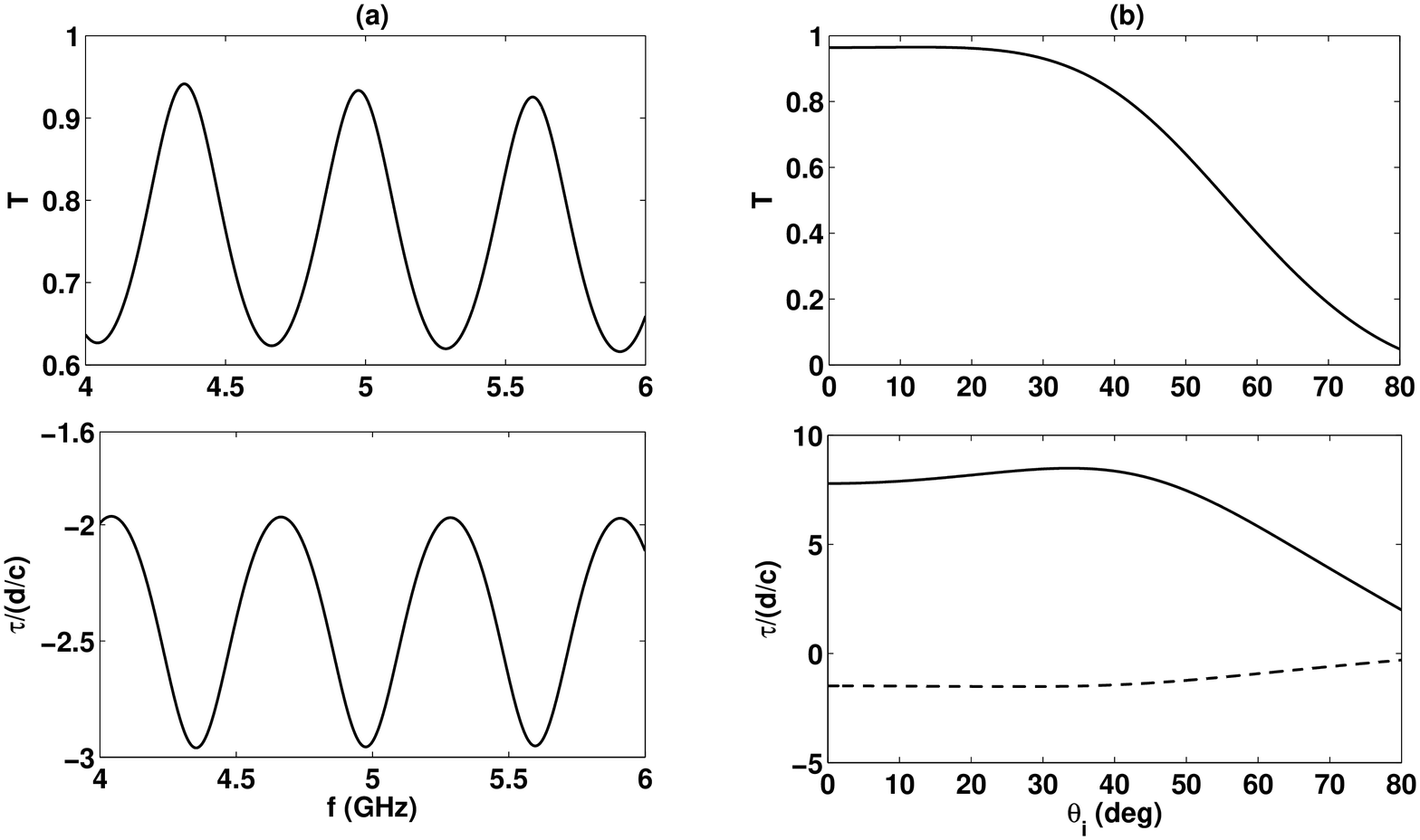}\textit{}
\caption{(a) Intensity transmission coefficient $T$  (top) and the normalized delay $\tau/(d/c)$ (bottom) for a FP cavity as functions of frequency; (b) the same as functions of angle of incidence $\theta_i$ for $f=5$ GHz, $\Gamma/\omega_0 =10^{-3}$ and $d=10$ cm. The bottom right panel compares the normalized delays with (solid) and without (dashed) material dispersion.}\label{fig:2}
\end{figure}
material, embedded in air,  the FP resonances are imprinted on this background leading to large delays at these resonances. Now we demonstrate how suppression of the material dispersion can lead to completely erroneous conclusions. Consider the same FP slab, but now with frequency independent $ \epsilon = \epsilon_{f=5{\rm GHz}} $, $ \mu = \mu_{f=5{\rm GHz}}$. For reference we have chosen the material parameters at an intermediate frequency, namely, $f = 5 $ GHz. The Airy resonances of this FP cavity are shown in the top panel of figure \ref{fig:2}a. The bottom panel shows the corresponding normalized delay. As can be seen from this figure, just the structural dispersion of the slab for the said parameters leads to negative delay characteristics, which is just the opposite of the actual situation. The same results can be verified from the angle scan of the transmission and the normalized delay at a particular frequency of incident light. In figure \ref{fig:2}b we show the dependence of the intensity transmission $T$ and the normalized phase time $\tau/(d/c)$ as functions of the angle of incidence $\theta_i$ at $f=5$ GHz.  For each angle of incidence we evaluated the complex amplitude transmission coefficients at two infinitesimally close frequencies $\omega \pm \Delta\omega$ and extracted the corresponding phases. The ratio of the differences in the phases to that of the frequencies lead to the phase time at $\omega$. Results for the cases with (solid line) or without (dashed line) intrinsic dispersion are shown in the lower panel of figure \ref{fig:2}b. Material dispersion was suppressed by assuming $\epsilon(\omega \pm \Delta\omega) = \epsilon(\omega) $, $\mu(\omega \pm \Delta\omega) = \mu(\omega) $. Again it is clear that neglect of dispersion can lead to the incorrect result of negative delay.
\begin{figure}[h]
\centering
\includegraphics[width=7cm]{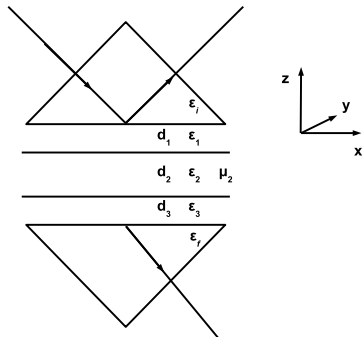}
\caption{Schematics of the layered structure with central NIM layer sandwiched between two spacer layers and high index prisms. All other materials except for the NIM are assumed to be non-magnetic.}\label{fig:3}
\end{figure}
\section{Delay in resonant tunnelling mediated by plasmon-like modes}
In this section we consider a structure which can lead to resonant tunnelling via the excitation of the surface (guided) modes. The structure is assumed to be symmetric ($\epsilon_i=\epsilon_f, ~ \epsilon_1=\epsilon_3, ~ d_1=d_3$) with the central layer made up of NIM (see figure \ref{fig:3}). In order to establish a proper connection with the current experimental activities, and also to test their feasibility with our current proposal, we pick the NIM parameters like $\epsilon(\omega)$ and $\mu(\omega)$ from the experiment of Dolling \etal \cite{dolling2006}. As mentioned earlier, despite the huge volume of literature on novel metamaterials, very few give the complete dispersion data (real and imaginary parts of both $\epsilon$ and $\mu$) for the NIM \cite{valentine2008,zhang2005,kildishev2006}. Most of these data are extracted from transmission studies. The work of Dolling \etal reported a figure of merit of about 3 at $\lambda \sim 1.4 \mu$m, where $Re(n)=-1$, much needed for perfect lensing applications. We used the data digitized from their results for $\epsilon$ and $\mu$ with two significant changes. It is understood that a thin layer of (120nm in their experiment) metamaterial is highly anisotropic. In our calculations we assumed the NIM material to be isotropic and homogeneous. Hopefully 3d near isotropic metamaterials will be realized in near future. The second approximation concerns the losses in the metamaterial. We later comment on how the actual losses of currently available metamaterials can wash out all the interesting effects reported here. Thus the other issue concerns the dire need of truly low-loss metamaterials. Instead of the actual experimentally observed losses, we assume a small loss of $i\gamma$ at the working frequency $\overline{\omega}$ for both permittivity and permeability. Thus at the specified frequency for the NIM we write
\begin{eqnarray}
\eqalign{
\epsilon_2(\overline{\omega}) &= Re(\epsilon_d(\overline{\omega})) + i\gamma,   \\
\mu_2(\overline{\omega}) &= Re(\mu_d(\overline{\omega})) + i\gamma,  
}\label{eq:epsmugamma} 
\end{eqnarray}
where $\epsilon_d(\overline{\omega})$ and $\mu_d(\overline{\omega})$ are taken from the work of Dolling \etal \cite{dolling2006}. The motivation for introducing the small damping is to understand, at least qualitatively, the effect of damping on the resonant tunnelling through a NIM guide. Henceforth, in this and the next section, we consider only the low-loss NIM guide with properties given by (\ref{eq:epsmugamma}) excited by a TE-polarized light.  
\begin{figure}[b]\centering
\includegraphics[width=12cm]{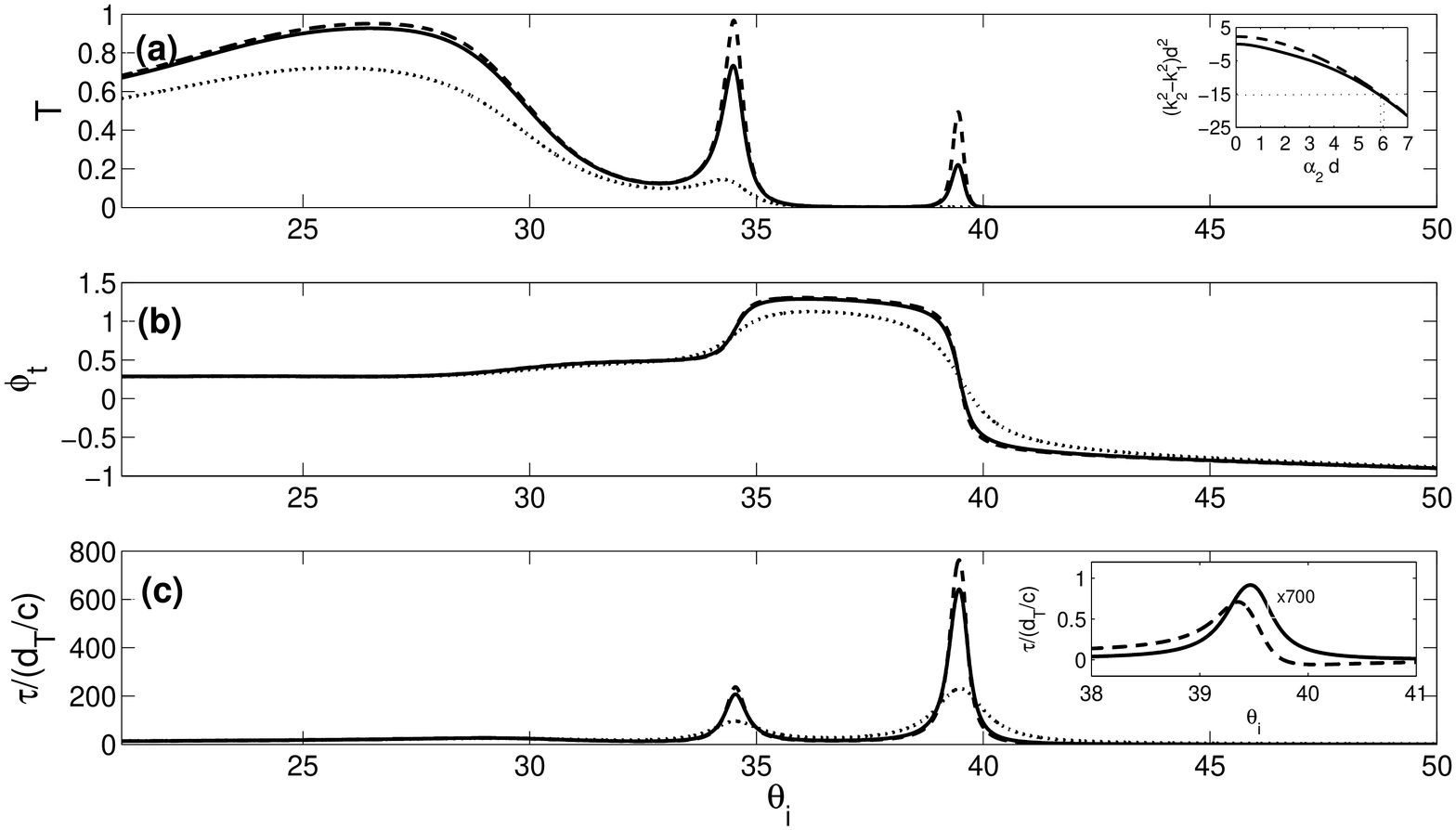}
\caption{(a) Intensity transmission coefficient T (b) phase of transmission (in units of $\pi$) and (c) normalized delay as a function of the angle of incidence. The parameters used are $ \lambda = 1.425\mu m, \epsilon_i = \epsilon_f = 6.145, \epsilon_1 = \epsilon_3 = 2.25, d_1 = d_3 = 1 \mu m, d_2 = 2 \mu m$ for $\gamma = $ 0.0001 (dashed), 0.001 (solid) and 0.01 (dotted). (Figure \ref{fig:4}a inset) Plot of $(k_2^2-k_1^2)d^2$ as a function of $\alpha_2 d$ as in the work of Wu \etal \cite{kong2003}. (Figure \ref{fig:4}c inset) Normalized delays with (solid lines, values reduced by a factor of 700) and without (dashed lines) material dispersion for $\gamma=0.001$. }\label{fig:4}
\end{figure}
\par
In order to have the plasmon like modes of the NIM guide we choose the spacer layer refractive index $n_1$ larger than $|Re(n_2)|$ at the working wavelength. In particular, we choose the following system parameters: $ \lambda = 2\pi\overline{\omega}/c = 1.425\mu m, \epsilon_i = 6.145, \epsilon_1 = \epsilon_3 = 2.25, d_1 = d_3 = 1 \mu m, d_2 = 2 \mu m$. The thickness of the spacer layers was chosen so as to optimize the excitation of a given mode of a NIM guide. The results for the intensity transmission ($T$), phase of transmission ($\phi_t$) and corresponding normalized delay ($\tau/(d_T/c)$, $d_T=d_1+d_2+d_3$) as functions of the angle of incidence are shown in figure \ref{fig:4} a, b and c, respectively. The dashed, dotted and solid curves are for three different values of $\gamma$, namely, $\gamma = $ 0.0001, 0.001 and 0.01, respectively. The nature of the peaks in figure \ref{fig:4} a and c can be easily understood if one recalls the critical angles for the relevant interfaces at the specified wavelength. For the prism/silica interface the critical angle is about 37.24$^\circ$ while that for the prism/NIM interface it is  about 35.25$^\circ$. For $\theta_i>37.24^\circ$ waves are evanescent in silica as well as the NIM slab. Thus the peak occurring at $\theta \sim 39^\circ$ can be associated with the surface-plasmon-like mode of the NIM guide. Such a mode is the dual of the TM-polarized surface plasmon polariton in a metal film \cite{kong2003} and have been discussed in detail in \cite{kivshar2003,kong2003} and used even for QED applications \cite{xu2009}.
\par
Following Wu \etal \cite{kong2003} we now determine the symmetry of the mode occuring at $\theta_i=39.45^\circ$. For this purpose we consider a symmetric NIM guide with width $d_2=2\mu m$ embedded in silica (with $\epsilon=2.25$). For such a guide with $\mu_1/\mu_2=-0.7487$ the symmetric (cosh-type) and anti-symmetric (sinh-type) mode dispersion curves are shown in the inset of figure \ref{fig:4}a. From the parameters used in figure \ref{fig:4}a, $(k_2^2-k_1^2)d^2=-15.7594$ (horizontal line in the inset of figure \ref{fig:4}a). We obtain two values of $\alpha_2 d=(k_x^2-k_2^2)^{1/2} d$ corresponding to the symmetric (solid curve) and anti-symmetric (dashed curve) modes. For the symmetric mode $\alpha_2 d \sim 5.949$ (vertical line in the inset of figure \ref{fig:4}a). The excitation angle corresponding to this value of $\alpha_2 d$ is 39.65$^\circ$ and is closer to the value of $\theta_i$ from the RT data. Thus the resonance at $\theta_i=39.45^\circ$ in figure \ref{fig:4}a is recognized as the symmetric plasmon-like mode. The peak close to $\theta_i=34.5^\circ$ corresponds to one of the FP modes of the layered medium, since waves are propagating both in silica as well as in the NIM layer. It is clear from figure \ref{fig:4}b that the phase of the transmission coefficient undergoes qualitatively different (positive and negative) changes as one sweeps the angles through the FP or the plasmon-like resonances. The oscillatory (plasmonic) modes are associated with positive (negative) jumps in phase. However, both lead to large delays. It is also clear how increasing damping gradually erases the RT features. For values of NIM losses as in \cite{dolling2006}, the transmission peak due to the `plasmonic' mode gets washed completely though the enhanced delay properties survive. Of course, there is no use of a delayed signal if its peak amplitude is vanishingly small. FP modes and the associated delay still survive to some extent even for large losses. One other important aspect that should be noted from figure \ref{fig:4} is that the plasmon-like modes can lead to much larger delays than the FP modes, due to the large quality factors associated with such modes.
\section{Delay in resonant tunnelling mediated by guided modes}
In this section, we consider again the RT configuration (see figure \ref{fig:3}), albeit with two changes. We choose a spacer layer with $\epsilon_1=\epsilon_3=1.0$, so that  $n_1< Re(n_2)$  and the guided modes could be excited.  In other words, the spacer layer is chosen to be optically rarer than the NIM. The other change is in the width of the spacer layers, namely, $d_1=d_3=0.25\mu m$, in order to optimize the excitation of the modes. The critical angles for the prism/air interface is 23.79$^\circ$ and prism/NIM interface (at $\lambda=1.425\mu m$) is 35.25$^\circ$. For angles greater than 23.79$^\circ$ and less than 35.25$^\circ$, the waves are evanescent in the spacer layers, while they are propagating in the NIM layer. Thus guided modes can be excited in the NIM layer only for this range of angles. Any peak observed in the transmission profile in this range can be recognized as due to the excitation of these guided modes.
\begin{figure}\centering
\includegraphics[width=15cm]{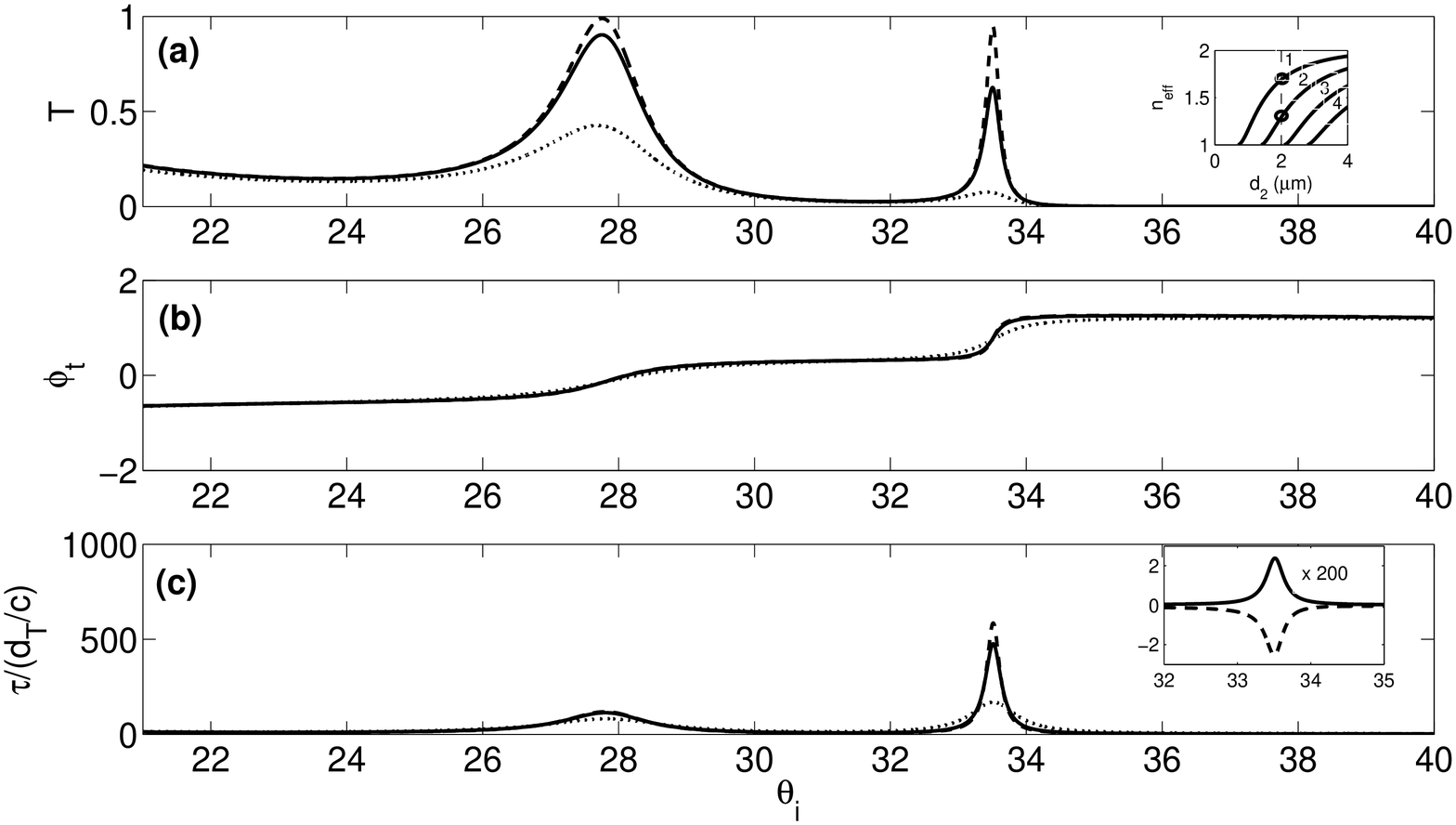}
\caption{Same as in figure \ref{fig:4} except that $\epsilon_1 = \epsilon_3 = 1, ~\mu_1=\mu_3=1, ~d_1 = d_3 = 0.25 \mu m$. Inset in figure \ref{fig:5}a shows the effective index of the modes as a function of guide width. Inset in figure \ref{fig:5}c shows the normalized delays with (solid lines, values reduced by a factor of 200) and without (dashed lines) material dispersion for $\gamma=0.001$. }\label{fig:5}
\end{figure}
\par
We show the intensity transmission $T$, phase of transmission $\phi_t$ and the normalized delay $\tau/(d_T/c)$ as functions of the angle of incidence $\theta_i$ in figures \ref{fig:5}a, b and c, respectively. The dashed, solid and dotted curves correspond to $\gamma=$ 0.0001, 0.001 and 0.01, respectively. It is clear from figure \ref{fig:5} that two guided modes corresponding to $\theta_i=$ 27.76$^\circ$ and 33.51$^\circ$, respectively, are supported in the NIM layer. In the inset of figure \ref{fig:5}a we plot the effective index, $n_{\rm eff}=k_x/k_0 $, of the modes of a lossless NIM guide in air as a function of the guide width. Different curves are labeled by the corresponding mode numbers. Recall that for a NIM guide the $m=0$ mode does not exist and the lowest allowed mode ($m=1$) has a cutoff (also noted in \cite{kivshar2003,kong2003}). From the inset we confirm that for $d_2=2\mu m$ (vertical dashed line in the inset), only two guided modes (marked with circles) are supported. The excitation angles ($\theta_i$) for these values of $n_{\rm eff}$ are 31.61$^\circ$ (for $m=1$) and 27.28$^\circ$ (for $m=2$), respectively, and they match well with the RT data. The mismatch can be  attributed to the fact that losses were ignored. The loading of the guide with the prism in the RT configuration is another important source of this mismatch. The plot of the phase of transmission (figure \ref{fig:5}b) shows positive jumps at the location of the guided mode resonances (compare with figure \ref{fig:4}b). The plot of normalized delay (figure \ref{fig:5}c) shows positive delays for both the guided modes. The intensity of transmission gets considerably weakened with increased damping though the delay characteristic is retained. In the inset of figure \ref{fig:5}c we plot the normalized delays for $\gamma=0.001$ with (solid line, values reduced by a factor of 200) and without (dashed line) including the effect of the material dispersion. We note again that consideration of material dispersion removes the erroneous negative delay features (obtained by disregarding the material dispersion) of NIM. Finally, a comparison of figures \ref{fig:5}c and \ref{fig:4}c reveal that the delay with the plasmonic mode is usually higher than that with the guided mode. This is perhaps due to the tighter confinement of the field and the larger quality factor of the plasmonic mode.
\section{Effects of losses in the framework of a causal response}
The treatment in the previous sections (3 and 4) at a given frequency was simple and losses were introduced just to demonstrate their devastating effects on resonant tunnelling features. The realistic treatment of losses over the whole spectral domain is much more complicated. In order to have a causal response, one must have susceptibilities analytic in the upper half complex plane, which is a direct consequence of the fact that response functions vanish for negative arguments. Moreover, the real and imaginary parts of the susceptibility are related by the Kramers-Kronig relation. Recent analysis of causality in the context of NIM dispersion sheds much insight on the underlying phenomena (with or without additional gain) \cite{stockman2007, kinsler2008prl, nistad2008prl}. These studies are based on the analytic properties of $n^2(\omega)=\epsilon(\omega)\mu(\omega)$. 
\begin{figure}[h]\centering
\includegraphics[width=15cm]{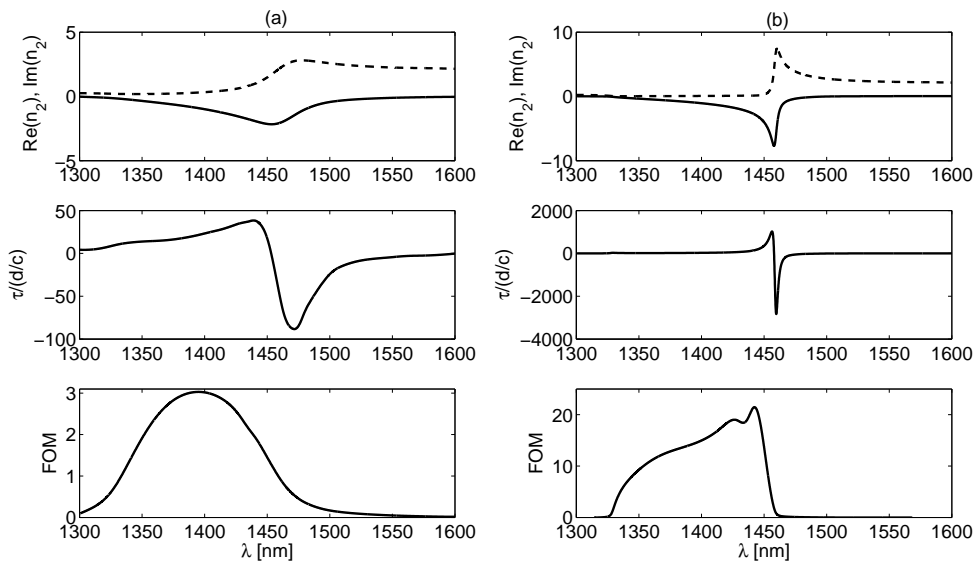}
\caption{The real (solid) and imaginary (dashed) parts of refractive index (top row), the normalized delay (middle row) and FOM (bottom row) as a function of $\lambda$. The left top panel shows the fitted curves for digitized data from Dolling \etal \cite{dolling2006}. The right panel shows the corresponding curves for improved magnetic resonance ($f=10$).}\label{fig:6}
\end{figure}
A compensation or reduction of material losses narrows down the regime over which the left-handed behaviour is observed. On the other hand, as noted in the preceeding sections, increasing loss washes out the interesting features associated with NIM guides. Attainment of a lossless NIM thus reaches an impasse. There should thus be a trade-off between the material loss and the regime of negative index behaviour for practical metamaterials. Magnetic losses are known to play a dominant role in controlling the negative index behaviour of metamaterials. In this section we analyze a material with Lorentz type magnetic permeability given by \cite{zhang2005oe}
\begin{equation}
\mu (\omega) = \mu_\infty - \frac{F \omega_0^2}{ \omega^2 - \omega_0^2+ i\gamma_m \omega},\label{eq:mucausal}
\end{equation}
where $\mu_\infty$, $\omega_0$, $F$ and  $\gamma_m$ have their usual meanings. Note that the experimental data of Dolling \etal \cite{dolling2006} can be fitted to (\ref{eq:mucausal}) with the following values of the parameters $\mu_\infty=0.6$, $\lambda_0=2\pi c/\omega_0$=1.459 $\mu m$, $\gamma_m/\omega_0=0.028$ and $F=4.425 (\gamma_m/\omega_0)$. We used the digitized data for the permittivity. The dependence for the said parameter values is shown in the top panel of figure \ref{fig:6}a, while the middle and the bottom panels show the corresponding normalized delay in bulk sample and the FOM, respectively. It was mentioned earlier that in a guide with such material parameters, the RT features are completely washed out. Hence we probe the case of an NIM with improved magnetic response, whereby we scale down the damping $\gamma_m$ by a factor $f_\gamma$ with simultaneous scaling up of the oscillator strength $F$ by $f_o$. The corresponding improved response for $f_o=f_\gamma=10$ is shown in figure \ref{fig:6}b. 
\begin{figure}[b]\centering
\includegraphics[width=15cm]{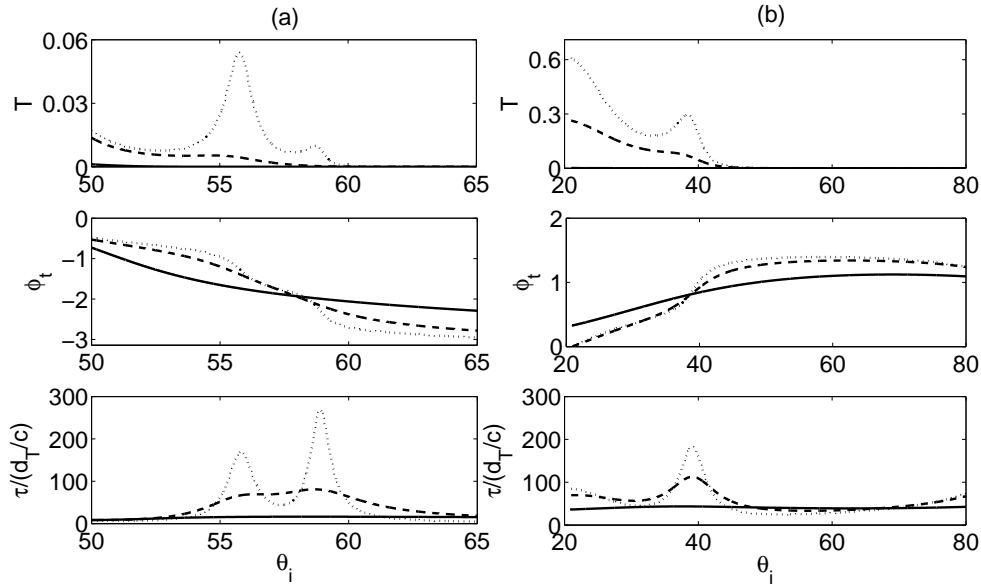}
\caption{Intensity transmission coefficient $T$ (top row), phase of transmission $\phi_t$ in units of $\pi$ (middle row) and the time delays (bottom row) as functions of the angle of incidence $\theta_i$ for the RT structure supporting plasmon-like modes (a) and guided modes (b). Parameters chosen for panel (a) are $\epsilon_1=\epsilon_3=4$, $d_1 = d_3 = 1 \mu m$, $d_2 = 0.5 \mu m$ and for panel (b) are $\epsilon_1=\epsilon_3=1$, $d_1 = d_3 = 0.05 \mu m$, $d_2 = 1 \mu m$. The common parameters are $\epsilon_i=\epsilon_f=6.145$ and $f=$1 (solid), 10 (dashed), 100 (dotted).}\label{fig:7}
\end{figure}
\par
We studied the transmission and pulse delay through a RT geometry for $f_o=f_\gamma=10, 100$ and compared it with the material of Dolling \etal \cite{dolling2006} ($f_o=f_\gamma=1$). Delay in RT mediated by plasmon-like modes were studied in a structure with $d_1=d_3=1\mu m$, $\epsilon_1=\epsilon_3=4$ and $d_2=0.5\mu m$ at $\lambda=1.425\mu m$ and the results are shown in figure \ref{fig:7}a. The critical angles for the prism/NIM interface is $44.42^\circ$ and the prism/spacer layer interface is $53.79^\circ$. We show that plasmon-like
 modes can be supported in this structure with large time delays for improved magnetic responses. These features disappear for the experimental data of Dolling \etal \cite{dolling2006}. A study of time delays mediated by guided modes with parameters $d_1=d_3=0.05\mu m$, $\epsilon_1=\epsilon_3=1$ and $d_2=1.0\mu m$ at $\lambda=1.425\mu m$ (figure \ref{fig:7}b) shows that the supported guided mode shows enhanced transmission and delay for improved magnetic responses ($f_o=f_\gamma=10,100$) whereas the data corresponding to Dolling \etal \cite{dolling2006} ($f_o=f_\gamma=1$) barely shows any transmission or delay features. Thus it should be possible to have enhanced delay characteristics with materials having improved magnetic responses.
\section{Conclusions}
We studied pulse delay through a layered medium containing a NIM layer. We considered three specific cases, namely, a generic NIM, a realistic, recently reported NIM with the assumption of low losses and NIM with improved magnetic response. We also assumed the NIM to be homogeneous. We showed that large delays in transmission, mediated by the modes of the NIM guide, are achievable exploiting a resonant tunneling geometry only for very low loss structures. Both guided and plasmon-like modes were exploited to this end. We probed the role of structural and intrinsic NIM dispersion in this delay and stressed the need to retain both for its correct assessment. We also commented on how the large damping of contemporary metamaterials can destroy the RT transmission features. We probed a NIM with better magnetic response to show the reemergence of these features. These calculations are carried out in the framework of a full causal theory. We thus ingeminated Veselago's observation about the importance and necessity of retaining intrinsic NIM dispersion in all relevant problems. We also emphasized the need for truly low-loss metamaterials for the realization of such compact delay devices.

The authors are thankful to the Department of Science and Technology, Government of India, for supporting this work. SDG also thanks Girish S Agarwal for interesting discussions.


\end{document}